\documentclass[10pt]{article}
\usepackage{multicol}
\usepackage{graphicx}
\usepackage{amsmath}
\usepackage[a4paper]{geometry}
\usepackage{verbatim}
\usepackage{hyperref}

\begin{document}

\begin{center}

{\Large \bf A new description of transverse momentum spectra of
identified particles produced in proton-proton collisions at high
energies}

\vskip.75cm

Pei-Pin~Yang$^{1,2,}${\footnote{E-mail: peipinyangshanxi@163.com;
yangpeipin@qq.com}}, Fu-Hu~Liu$^{1,2,}${\footnote{Corresponding
author. E-mail: fuhuliu@163.com; fuhuliu@sxu.edu.cn}},
Raghunath~Sahoo$^{3,}${\footnote{E-mail: raghunath.phy@gmail.com;
Raghunath.Sahoo@cern.ch}}

\vskip.25cm

{\small\it $^1$Institute of Theoretical Physics \& State Key
Laboratory of Quantum Optics and Quantum Optics Devices,\\ Shanxi
University, Taiyuan, Shanxi 030006, People's Republic of China

$^2$Collaborative Innovation Center of Extreme Optics, Shanxi
University,\\ Taiyuan, Shanxi 030006, People's Republic of China

$^3$Discipline of Physics, School of Basic Sciences, Indian
Institute of Technology Indore,\\ Simrol, Indore 453552, Republic
of India}

\end{center}

\vskip.5cm

{\bf Abstract:} The transverse momentum spectra of identified
particles produced in high energy proton-proton ($p+p$) collisions
are empirically described by a new method with the framework of
participant quark model or the multisource model at the quark
level, in which the source itself is exactly the participant
quark. Each participant (constituent) quark contributes to the
transverse momentum spectrum, which is described by the TP-like
function, a revised Tsallis--Pareto-type function. The transverse
momentum spectrum of the hadron is the convolution of two or more
TP-like functions. For a lepton, the transverse momentum spectrum
is the convolution of two TP-like functions due to two participant
quarks, e.g. projectile and target quarks, taking part in the
collisions. A discussed theoretical approach seems to describe the
$p+p$ collisions data at center-of-mass energy $\sqrt{s}=200$ GeV,
2.76 TeV, and 13 TeV very well.

\vskip1.0cm

\begin{multicols}{2}

{\section{Introduction}}

As one of the ``first day" measurable quantities, the transverse
momentum ($p_{\rm T}$) spectra of various particles produced in
high energy proton-proton ($p+p$) (hadron-hadron), proton-nucleus
(hadron-nucleus), and nucleus-nucleus collisions are of special
importance because, it reveals about the excitation degree and
anisotropic collectivity in the produced systems. The distribution
range of $p_{\rm T}$ is generally very wide, from 0 to more than
100 GeV/$c$, which is collision energy dependent. In the very
low-, low-, high-, and very high-$p_{\rm T}$ regions~\cite{1}, the
shapes of $p_{\rm T}$ spectrum for given particles are possibly
different from each other. In some cases, the differences are very
large and the spectra show different empirical laws.

Generally, the spectrum in (very) low-$p_{\rm T}$ region is
contributed by (resonance decays or other) soft excitation
process. The spectrum in (very) high-$p_{\rm T}$ region is related
to (very) hard scattering process (pQCD). There is no clear
boundary in $p_{\rm T}$ to separate soft and hard processes. At a
given collision energy, for different collision species, looking
into the spectral shape, a theoretical function that best fits to
the $p_{\rm T}$-spectra is usually chosen to extract information
like rapidity density, $dN/dy$, kinetic freeze-out temperature,
$T_{\rm kin}$ or $T_0$ and average radial flow velocity, $\langle
\beta_{\rm T}\rangle$ or $\beta_T$. The low-$p_{\rm T}$ region up
to $\sim2$--3 GeV/$c$ is well described by a Boltzmann--Gibbs
function, whereas the high-$p_{\rm T}$ part is dominated by a
power-law tail. It is interesting to note that there are many
different functions, sometimes motivated by experimental trend of
the data or sometimes theoretically, to have a proper spectral
description thereby leading to a physical picture. The widely used
functions are:
\begin{enumerate}
\item{An exponential function in $p_T$ or
$m_T$~\cite{starSyst,2a}:
\begin{equation}
\label{eq:spectra_fits:exppt} f(p_{\rm{T}}) = p_{\rm{T}}\times
A\times\left(e^{-p_{\rm{T}}/T}\right)\times
\frac{e^{m_0/T}}{T^{2}+Tm_0},
\end{equation}
\begin{equation}
\label{eq:spectra_fits:expmt} f(p_{\rm{T}}) = p_{\rm{T}}\times
A\times\left(e^{-m_{\rm{T}}/T}\right)\times\frac{e^{m_0/T}}{T^{2}+Tm_0}.
\end{equation}
Here, $A$ is the normalization constant, $T$ is the effective
temperature (thermal temperature and collective radial flow) and
$m_{\rm T} =\sqrt{p_{\rm T}^2+m_0^2}$ is the transverse mass, with
$m_0$ being the identified particle rest mass. } \item{A Boltzmann
distribution:
\begin{align}
\label{eq:spectra_fits:boltzmann} f(p_{\rm{T}}) & = p_{\rm{T}}
\times A\times m_{\rm{T}}\times\left(e^{-m_{\rm{T}}/T}\right) \nonumber\\
& \times\frac{e^{m_0/T}}{2T^{3}+2T^{2}m_0+Tm_0^{2}}.
\end{align}
} \item{Bose--Einstein/Fermi--Dirac distribution:
\begin{align}
\label{eq:spectra_fits:bose} f(p_{\rm{T}}) &= p_{\rm{T}}\times
A\times m_{\rm{T}}\times \frac{1}{e^{m_{\rm{T}}/T}\mp1} \nonumber\\
&\times\left(e^{m_0/T}\mp1\right),
\end{align}
} \item{Power-law or Hagedorn function~\cite{3}:
\begin{eqnarray}
 f(p_{\rm{T}})& = & p_{\rm{T}} \times A\times \left( 1 + \frac{p_{\rm T}}{p_0}\right)^{-n}
\nonumber\\
 & \longrightarrow&
  \left\{
 \begin{array}{l}
  \exp\left(-\frac{n p_{\rm T}}{p_0}\right),\quad \, \, \, {\rm for}\ p_{\rm T} \to 0, \smallskip\\
  \left(\frac{p_0}{p_{\rm T}}\right)^{n},\qquad \qquad{\rm for}\ p_{\rm T} \to \infty,
 \end{array}
 \right .
 \label{eq2}
\end{eqnarray}
where $p_0$ and $n$ are fitting parameters. This becomes a purely
exponential function for small $p_{\rm T}$ and a purely power-law
function for large $p_{\rm T}$ values. } \item{Tsallis--L\'{e}vy
\cite{STAR_strange_pp_2007,6} or Tsallis--Pareto-type
function~\cite{6,7,7a,7b}:
\begin{align}
\label{eq:spectra_fits:levy} f(p_{\rm{T}}) &=
p_{\rm{T}}\times\frac{A(n-1)(n-2)}{nT\left[nT+m_0(n-2)\right]}
\nonumber\\
&\times \left(1+\frac{m_{\rm{T}}-m_0}{nT}\right)^{-n}.
\end{align}
}
\end{enumerate}

Note here that a multiplicative pre-factor of $p_{\rm T}$ in the
above functions are used assuming that the $p_{\rm T}$-spectra do
not have a $p_{\rm{T}}$ factor in the denominator (see the
expression for the invariant yield) and all the functions are
normalized so that the integral of the functions provides the
value of ``$A$". When the first three functions describe the
$p_{\rm T}$-spectra up to a low-$p_{\rm T}$ around 2--3 GeV/$c$,
the fourth function i.e. the power-law describes the high-$p_{\rm
T}$ part of the spectrum. The last two functions (power-law or
Hagedorn function and Tsallis--L\'{e}vy or Tsallis--Pareto-type
function), which are more empirical in nature, lack microscopic
picture, however, describe wide variety of identified particle
spectra. The Tsallis distribution function, while describing the
spectra in $p+p$ collisions \cite{thakur}, has brought up the
concept of non-extensive entropy, contrary to the low-$p_{\rm T}$
domain pointing to an equilibrated system usually described by
Boltzmann-Gibbs extensive entropy. In addition, identified
particle spectra are successfully explained in heavy-ion
collisions with the inclusion of radial flow in a Tsallis Blast
Wave description \cite{zhangbu}.

The two behaviors in (very) low- and (very) high-$p_{\rm T}$
regions are difficult to fit simultaneously by a simple
probability density function. Instead, one can use a two-component
function~\cite{2}, the first component $f_1(p_{\rm T})$ is for the
(very) low-$p_{\rm T}$ region and the second component $f_2(p_{\rm
T})$ is for the (very) high-$p_{\rm T}$ region, to superpose a new
function $f(p_{\rm T})$ to fit the $p_{\rm T}$ spectra. There are
two forms of superpositions, $f(p_{\rm T})=kf_1(p_{\rm
T})+(1-k)f_2(p_{\rm T})$ or $f(p_{\rm T})=A_1\theta(p_1-p_{\rm
T})f_1(p_{\rm T})+A_2\theta(p_{\rm T}-p_1)f_2(p_{\rm
T})$~\cite{3,4,5}, where $k$ denotes the contribution fraction of
the first component, $A_1$ and $A_2$ are constants which make the
two components equal to each other at $p_{\rm T}=p_1$, and
$\theta(x)$ is the usual step function which satisfies
$\theta(x)=0$ if $x<0$ and $\theta(x)=1$ if $x\geq0$.

It is known that there are correlations in determining parameters
in the two components in the first superposition~\cite{4}. There
is possibly a non-smooth interlinkage at $p_{\rm T}=p_1$ between
the two components in the second superposition~\cite{5}. We do not
expect these two issues. To avoid the correlations and non-smooth
interlinkage, we hope to use a new function to fit simultaneously
the spectra in the whole $p_{\rm T}$ region for various particles.
After sounding many functions out, a Tsallis--Pareto-type
function~\cite{6,7,7a,7b} which empirically describes both the
low-$p_{\rm T}$ exponential and the high-$p_{\rm T}$
power-law~\cite{8,9,10,11} is the closest to our target, though
the Tsallis--Pareto-type function is needed to revise its form in
some cases.

In this work, to describe the spectra in the whole $p_{\rm T}$
range which includes (very) low and (very) high $p_{\rm T}$
regions, the Tsallis--Pareto-type function is empirically revised
by a simple method. To describe the spectra in the whole $p_{\rm
T}$ range as accurately as possible, the contribution of
participant quark to the spectrum is also empirically taken to be
the revised Tsallis--Pareto-type (TP-like) function with another
set of parameters. Then, the $p_{\rm T}$ distribution of given
particles is a convolution of a few TP-like functions. To describe
the spectra of identified particles in the whole $p_{\rm T}$
range, both the TP-like function and the convolution of a few
TP-like functions are used to fit the data measured in $p+p$
collisions at center-of-mass energy $\sqrt{s}=200$
GeV~\cite{12,13,14,15,16}, 2.76
TeV~\cite{17,18,19,20,21,22,23,24,25}, and 13
TeV~\cite{26,27,28,29,30,31,32} by different collaborations.

The remainder of this paper is structured as follows. The
formalism and method are described in Section 2. Results and
discussion are given in Section 3. In Section 4, we summarize our
main observations and conclusions.
\\

{\section{Formalism and method}}

According to refs.~\cite{6,7,7a,7b}, the Tsallis--Pareto-type
function which empirically describes both the low-$p_{\rm T}$
exponential and the high-$p_T$ power-law can be simplified as
presented in~\cite{8,9,10,11},
\begin{align}
f(p_{\rm T})= C\times p_{\rm T}\times \left(1+\frac{\sqrt{p_{\rm
T}^2+m_0^2}-m_0}{nT}\right)^{-n}
\end{align}
in terms of $p_{\rm T}$ probability density function, where the
parameter $T$ describes the excitation degree of the considered
source, the parameter $n$ describes the degree of non-equilibrium
of the considered source, and $C$ is the normalization constant
which depends on $T$, $n$, and $m_0$. Equation (7) is in fact an
improvement of Eq. (6).

As an empirical formula, the Tsallis--Pareto-type function is
successful in the description of $p_{\rm T}$ spectra in many
cases. However, our exploratory analysis shows that Eq. (7) in
some cases is not accurate in describing the spectra in the whole
$p_{\rm T}$ range. In particular, Eq. (7) is not flexible enough
to describe the spectra in very low-$p_{\rm T}$ region, which is
contributed by the resonance decays. We would like to revise
empirically Eq. (7) by adding a power index $a_0$ on $p_{\rm T}$.
After the revision, we have
\begin{align}
f(p_{\rm T})= C\times p_{\rm T}^{a_0}\times \left( 1+
\frac{\sqrt{p_{\rm T}^2+m_0^2}-m_0}{nT} \right)^{-n},
\end{align}
where $C$ is the normalization constant which is different from
that in Eq. (7). To be convenient, the two normalization constants
in Eqs. (7) and (8) are denoted by the same symbol $C$. Eq. (8)
can be used to fit the spectra in the whole $p_{\rm T}$ range. The
revised Tsallis--Pareto-type function [Eq. (8)] is called the
TP-like function by us.

It should be noted that the index $a_0$ is a quantity with
non-dimension. Because of the introduction of $a_0$, the dimension
of $p_{\rm T}^{a_0}$ is $({\rm GeV}/c)^{a_0}$. The dimension of
$p_{\rm T}^{a_0}$ does not affect the dimension $({\rm
GeV}/c)^{-1}$ of $f(p_{\rm T})$. In fact, to fit the dimension of
$f(p_{\rm T})$, the dimension of the product $Cp_{\rm T}^{a_0}$ is
limited to be $({\rm GeV}/c)^{-1}$. That is to say, the dimension
of $p_{\rm T}^{a_0}$ is combined in the normalization constant so
that we can obtain the consistent dimension for both sides of the
equation. Due to the introduction of $a_0$, for the spectra in
very low-$p_{\rm T}$ region, not only the production of light
particles via resonance decay but also the decay or absorbtion
effect of heavy particles in hot and dense medium in participant
region can be described.

Our exploratory analysis shows that Eq. (8) is not accurate in
describing the spectra in the whole $p_T$ range, too, though it is
more accurate than Eq. (7). To obtain accurate results, the amount
or portion ($p_{ti}$) contributed by the $i$-th participant quark
to $p_T$ is assumed to obey
\begin{align}
f_i(p_{ti})= C_i\times p_{ti}^{a_0}\times \left( 1+
\frac{\sqrt{p_{ti}^2+m_{0i}^2}-m_{0i}}{nT} \right)^{-n},
\end{align}
where the subscript $i$ is used for the quantities related to the
participant quark $i$, and $m_{0i}$ is empirically the constituent
mass of the considered quark $i$. The value of $i$ can be 2 or 3
even 4 or 5 due to the number of participant (or constituent)
quarks. Eq. (9) is also the TP-like function with different mass
from Eq. (8).

It should be noted that $m_0$ in Eq. (8) is for a particle, and
$m_{0i}$ in Eq. (9) is for the quark $i$. For example, if we study
the $p_{\rm T}$ spectrum of protons, we have $m_0=0.938$ GeV/$c^2$
and $m_{01}=m_{02}=m_{03}=0.31$ GeV/$c^2$. In the case of studying
the $p_{\rm T}$ spectrum of photons, we have $m_0=0$ and
$m_{01}=m_{02}=0.31$ GeV/$c^2$ if we assume that two lightest
quarks take part in the collision with photon production.

There are two participant quarks to constitute usually mesons,
namely the quarks 1 and 2. The $p_{\rm T}$ spectra of mesons are
the convolution of two TP-like functions. We have
\begin{align}
f(p_{\rm T})&=\int_0^{p_{\rm T}} f_1(p_{t1})f_2(p_{\rm T}-p_{t1})dp_{t1} \nonumber\\
&=\int_0^{p_{\rm T}} f_2(p_{t2})f_1(p_{\rm T}-p_{t2})dp_{t2}
\end{align}
in which $f_1(p_{t1})f_2(p_{t2})$ is the probability for given
$p_{t1}$ and $p_{t2}$. The total probability considered various
$p_{t1}$ and $p_{t2}$ is given by Eq. (10) which is the
convolution of distributions of two independent
variables~\cite{32a,32b}. The upper limit $p_T$ is not a cutoff,
but the sum of $p_{t1}$ and $p_{t2}$, which is limited by the
physics. The lower limit 0 is also from the limitation related to
the underlying physics. No matter how many leptons are produced in
the process, two participant quarks are considered to contribute
to the $p_T$ spectrum of each lepton.

We would like to explain our treatment on Eq. (10) here. At least
three relations between particle $p_T$ and quark $p_{t1}$
($p_{t2}$) can be assumed. i) If we regard $p_{t1}$ ($p_{t2}$) as
the amount or portion contributed by the first (second)
participant quark to $p_T$, we have $p_T=p_{t1}+p_{t2}$. ii) If we
regard the vector $p_{t1}$ ($p_{t2}$) as the component contributed
by the first (second) participant quark to the vector $p_T$, we
have $p_T=\sqrt{p_{t1}^2+p_{t2}^2}$, where $p_{t1}$ is
perpendicular to $p_{t2}$. iii) In the second relation, it is not
necessary that all the components are perpendicular, then we have
\(p_T=\sqrt{p_{t1}^2+p_{t2}^2+2p_{t1}p_{t2}\cos|\phi_1-\phi_2|}\),
where \(\phi_1\) (\(\phi_2\)) is the azimuthal angle of the first
(second) participant quark. Different assumptions result in
different relations. Of course, the three \(p_{t1}\) (\(p_{t2}\))
in the three relations have different meanings, though the same
symbol is used. In our opinion, at present, it is hard to say
which relation is more correct. We need to test the three
relations by more experimental data.

In fact, all the three relations have still pending issues which
are needed further discussions. In the relation i), although
\(p_T\) can be considered as the contribution of two energy
sources: the first and second participant quarks that contribute
the amounts or portions \(p_{t1}\) and \(p_{p2}\) to \(p_T\)
respectively, the vector characteristic of transverse momentum is
not used. In the relation ii), as a vector, the transverse
momentum is considered by two components: \(p_{t1}\) and
\(p_{t2}\) which are contributed by the first and second
participant quarks respectively, though the origin of the third
component of meson momentum is not clear. In addition, although
the origin of three components of baryon momentum is clear, the
physics picture is not consistent to meson momentum. In the
relation iii), two more parameters \(\phi_1\) and \(\phi_2\) are
introduced, which is not our expectation.

This paper has used the relation i) and Eq. (10) which is based on
the probability theory~\cite{32a,32b,33}. However, in our recent
work~\cite{33a}, we have used the relation ii) and another
functional form which is based on the vector and probability
theory~\cite{32b,33}. We hope that we may use the relation iii) in
our future work by some limitations on \(\phi_1\) and \(\phi_2\).
The relation i) in terms of amount or portion is the same as or
similar to the relation for multiplicity or transverse energy
contributed by two sources~\cite{32a}. This similarity reflects
the law of universality existing in high energy
collisions~\cite{33b,33c,33d,33d1,33d2,33d3}. In fact, transverse
momentum, multiplicity, and transverse energy reflect the amount
of effective energy deposited in collisions~\cite{33e,33f}. The
effective energy through the participant quarks reflect the
similarity or universality, which is not related to the production
mechanisms for different particles. Then, different particles are
described by the same type of model (formula).

At the level of current knowledge, leptons have no further
structures. However, to produce a lepton in a common process, two
participant quarks, a projectile quark and a target quark, are
assumed to take part in the interactions. The $p_{\rm T}$ spectra
of leptons are in fact the convolution of two TP-like functions,
that is Eq. (10) in which $m_{01}$ and $m_{02}$ are empirically
the constituent mass of the lightest quark. To produce leptons in
a special process such as in $c\bar c \longrightarrow \mu^+\mu^-$,
$m_{01}$ ($m_{02}$) is the constituent mass of $c$ quark.

There are three participant quarks that constitute usually
baryons, namely the quarks 1, 2, and 3. The $p_{\rm T}$ spectra of
baryons are the convolution of three TP-like functions. We have
the convolution of the first two TP-like functions to be
\begin{align}
f_{12}(p_{t12})&=\int_0^{p_{t12}} f_1(p_{t1}) f_2(p_{t12}-p_{t1})
dp_{t1} \nonumber\\
&=\int_0^{p_{t12}} f_2(p_{t2}) f_1(p_{t12}-p_{t2}) dp_{t2}.
\end{align}
The convolution of the first two TP-like functions and the third
TP-like function is
\begin{align}
f(p_{\rm T})&=\int_0^{p_{\rm T}} f_{12}(p_{t12}) f_3(p_{\rm
T}-p_{t12}) dp_{t12}
\nonumber\\
&=\int_0^{p_{\rm T}} f_3(p_{t3}) f_{12}(p_{\rm T}-p_{t3}) dp_{t3}.
\end{align}

Equation (8) can fit approximately the spectra in the whole
$p_{\rm T}$ range for various particles at the particle level, in
which $m_0$ is the rest mass of the considered particle. In
principle, Eqs. (10) and (12) can fit the spectra in the whole
$p_T$ range for various particles at the quark level, in which
$m_{0i}$ is the constituent mass of the quark $i$. If Eq. (8) is
more suitable than Eq. (7), Eqs. (10) and (12) are the results of
the multisource model~\cite{34,35} at the quark level. In the
multisource model, one, two, or more sources are assumed to emit
particles due to different production mechanisms, source
temperatures and event samples. In a given event sample, the
particles with the same source temperature are assumed to emit
from the same source by the same production mechanism. We can also
call Eqs. (10) and (12) the results of participant quark model due
to the fact that they describe the contributions of participant
quarks.

It should be noted that, in principle, the three quarks should be
symmetric in the formula for the production of baryons. Indeed, in
Eqs. (11) and (12), the two momenta \(p_{t,1}\) and \(p_{t,2}\)
are symmetric, and the third momentum \(p_{t,3}\) is also
symmetric to the other two momenta. In fact, according to the rule
of the convolution of three functions, we may also convolute
firstly the last two functions, and then we may convolute the
result with the first function. Meanwhile, we may also convolute
firstly the first and third functions, and then we may convolute
the result with the second function. We realize that the final
result is not related to the order of convolution. The three
functions contributed by the three quarks are indeed symmetric.

We would like to explain the normalization constant in detail. As
a probability density function, $f(p_{\rm T})=(1/N) dN/dp_{\rm T}$
cannot be used to compare directly with the experimental data
presented in the literature in some cases, where $N$ denotes the
number of considered particles. Generally, the experimental data
are presented in forms of i) $dN/dp_{\rm T}$, ii) $d^2N/dydp_{\rm
T}$, and iii) $(1/2\pi p_{\rm T}) d^2N/dydp_{\rm T}=Ed^3N/dp^3$,
where $E$ ($p$) denotes the energy (momentum) of the considered
particle. One can use $N_0f(p_{\rm T})$, $N_0f(p_{\rm T})/dy$, and
$(1/2\pi p_{\rm T}) N_0f(p_{\rm T})/dy$ to fit them accordingly,
where $N_0$ denotes the normalization constant.

The data are usually in the form: i) $d\sigma/dp_{\rm T}$, ii)
$d^2\sigma/dydp_{\rm T}$, and iii) $(1/2\pi p_{\rm T})
d^2\sigma/dydp_{\rm T}=Ed^3\sigma/dp^3$, where $\sigma$ denotes
the cross-section. One can use $\sigma_0f(p_{\rm T})$,
$\sigma_0f(p_{\rm T})/dy$, and $(1/2\pi p_{\rm T})
\sigma_0f(p_{\rm T})/dy$ to fit them accordingly, where $\sigma_0$
denotes the normalization constant. The data presented in terms of
$m_{\rm T}$ can also be studied due to the conserved probability
density and the relation between $m_{\rm T}$ and $p_{\rm T}$. In
particular, $(1/2\pi p_{\rm T}) d^2\sigma/dydp_{\rm T} =(1/2\pi
m_{\rm T}) d^2\sigma/dydm_{\rm T}$, where $\sigma$ can be replaced
by $N$.

It should be noted that our treatment procedure means that the
parameters are fitted for each energy and rapidity bin separately.
This would limit the usefulness of the proposed parametrizations
somewhat. However, after obtaining the relations between
parameters and energy/rapidity, we can use the obtained fits to
predict $p_T$ distributions at other energies/rapidities where the
data are not available and the parameters are not fitted.
\\

\begin{figure*}[htbp]
\begin{center}
\includegraphics[width=15.0cm]{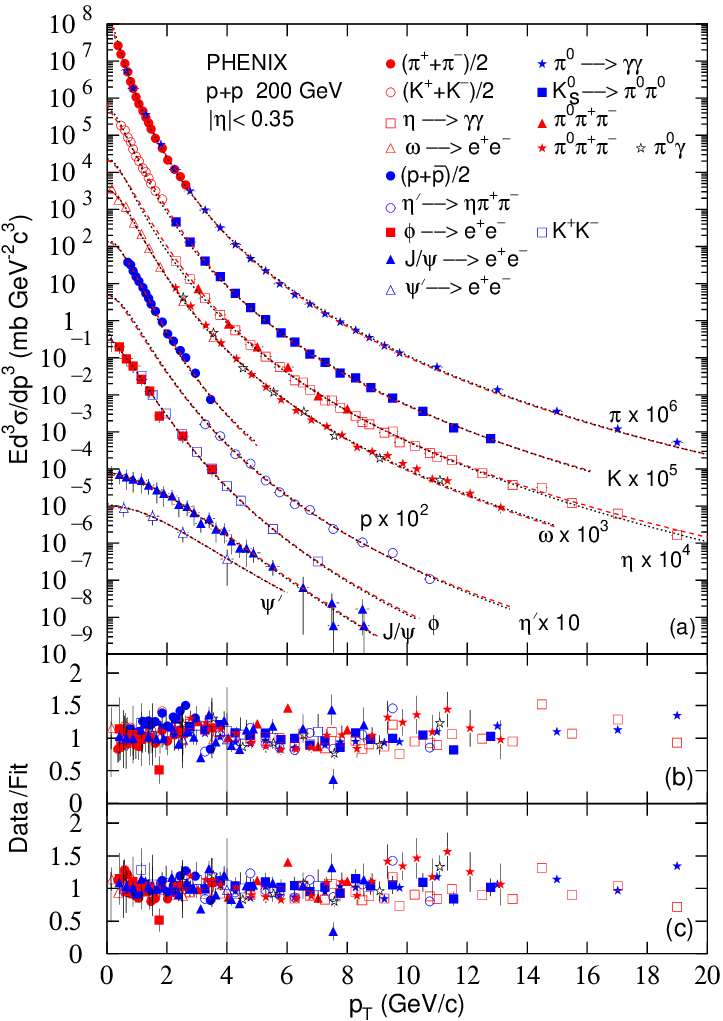}
\end{center}
{\small Fig. 1. (a) The invariant cross-sections of different
hadrons with given combinations and decay channels produced in
$p+p$ collisions at 200 GeV. Different symbols represent different
particles and their different decay channels in $|\eta|<0.35$
measured by the PHENIX Collaboration~\cite{12}, some of them are
scaled by different factors marked in the panel. The dotted and
dashed curves are our fitted results by using Eqs. (8) and (10) or
(12), respectively. (b) The ratio of data to fit obtained from Eq.
(8). (c) The ratio of data to fit obtained from Eq. (10) or (12).}
\end{figure*}

\begin{figure*}[htbp]
\begin{center}
\includegraphics[width=15.0cm]{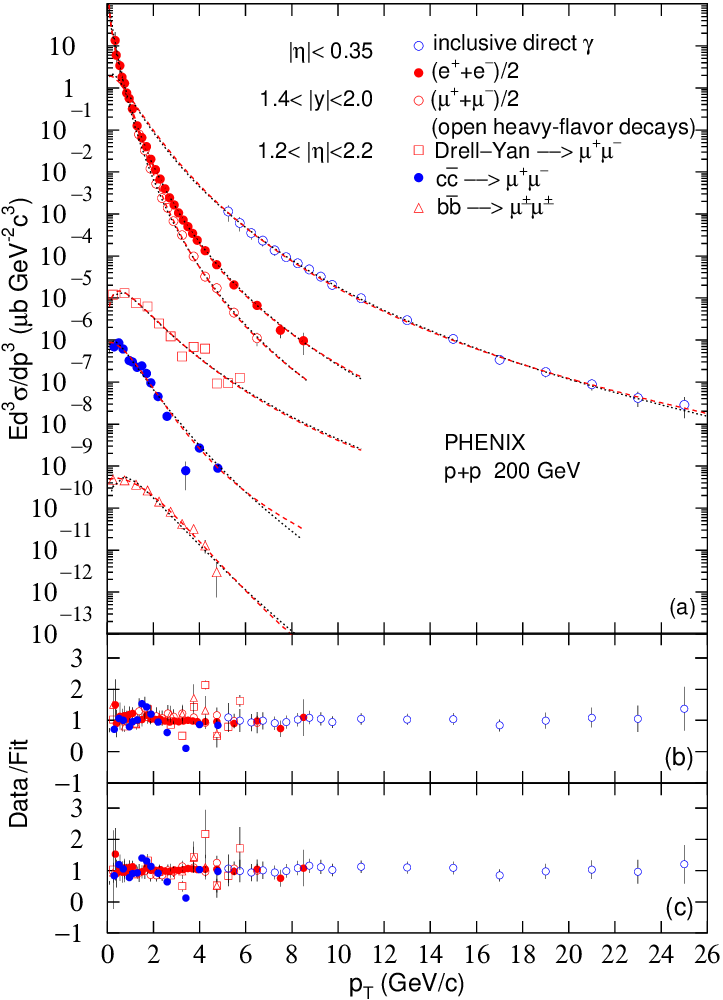}
\end{center}
{\small Fig. 2. (a) The invariant cross-sections of photons and
different leptons for a given combination of intermediate channel
for $p+p$ collisions at 200 GeV. Different symbols represent
different particles and their production channels in different
$\eta$ ranges measured by the PHENIX
Collaboration~\cite{13,14,15,16}. The dotted and dashed curves are
our fitted results by using Eqs. (8) and (10), respectively. (b)
The ratio of data to fit obtained from Eq. (8). (c) The ratio of
data to fit obtained from Eq. (10).}
\end{figure*}

\begin{figure*}[htbp]
\begin{center}
\includegraphics[width=15.0cm]{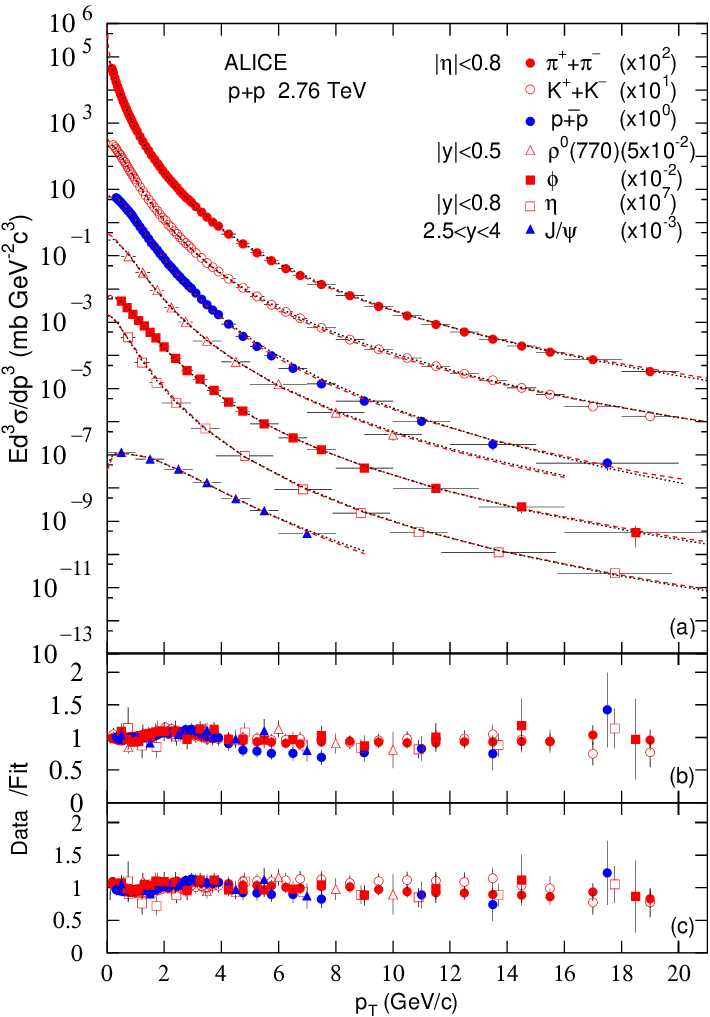}
\end{center}
{\small Fig. 3. (a) The invariant cross-sections of different
hadrons produced in $p+p$ collisions at 2.76 TeV. Different
symbols represent different particles in different $\eta$ or $y$
ranges measured by the ALICE Collaboration~\cite{17,18,19,20,21}
and scaled by different factors marked in the panel. The dotted
and dashed curves are our fitted results by using Eqs. (8) and
(10) or (12), respectively. (b) The ratio of data to fit obtained
from Eq. (8). (c) The ratio of data to fit obtained from Eq. (10)
or (12).}
\end{figure*}

\begin{figure*}[htbp]
\begin{center}
\includegraphics[width=15.0cm]{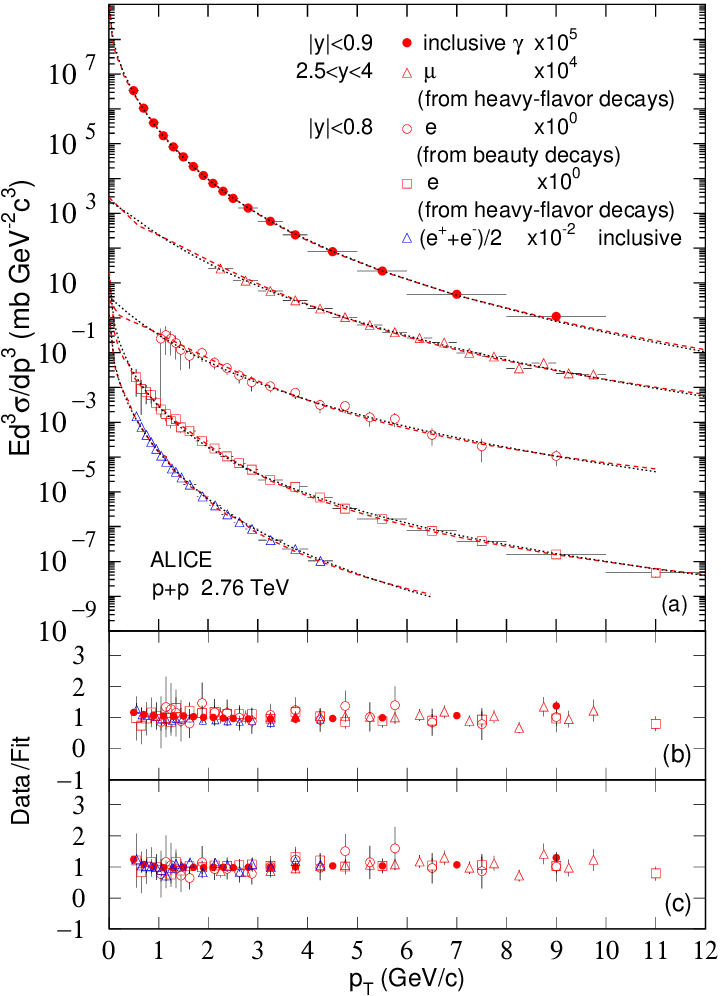}
\end{center}
{\small Fig. 4. (a) The invariant cross-sections of photons and
different leptons for a given combination of intermediate channel
for $p+p$ collisions at 2.76 TeV. Different symbols represent
different particles and their production channels in different $y$
ranges measured by the ALICE Collaboration~\cite{22,23,24,25} and
scaled by different factors marked in the panel. The dotted and
dashed curves are our fitted results by using Eqs. (8) and (10),
respectively. (b) The ratio of data to fit obtained from Eq. (8).
(c) The ratio of data to fit obtained from Eq. (10).}
\end{figure*}

\begin{figure*}[!htb]
\begin{center}
\includegraphics[width=15.0cm]{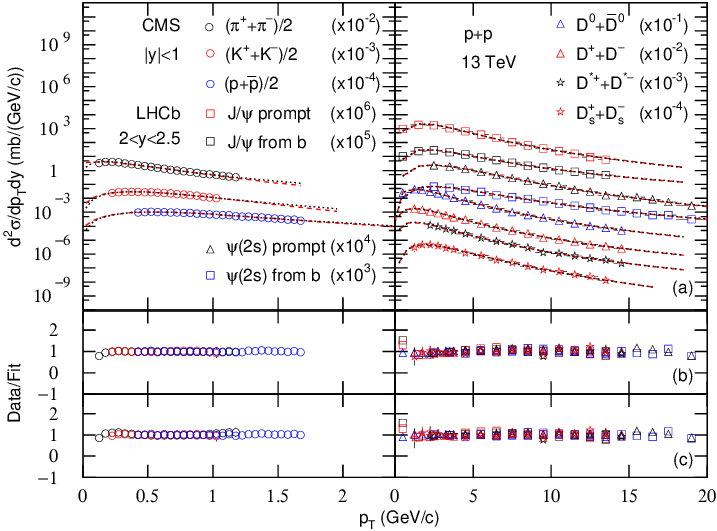}
\end{center}
{\small Fig. 5. (a) The invariant cross-sections of different
hadrons produced in $p+p$ collisions at 13 TeV. Different symbols
represent different particles in different $y$ ranges measured by
the CMS~\cite{26} and LHCb~\cite{27,28,29} Collaborations and
scaled by different factors marked in the panel. The dotted and
dashed curves are our fitted results by using Eqs. (8) and (10) or
(12), respectively. (b) The ratio of data to fit obtained from Eq.
(8). (c) The ratio of data to fit obtained from Eq. (10) or (12).}
\end{figure*}

\begin{figure*}[!htb]
\begin{center}
\includegraphics[width=10.0cm]{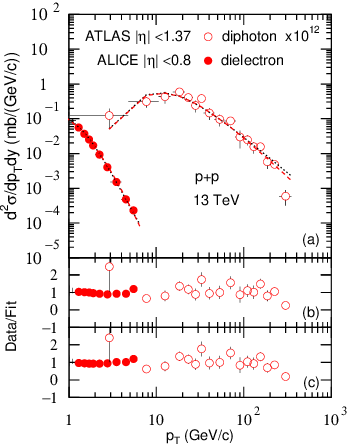}
\end{center}
{\small Fig. 6. (a) The invariant cross-sections of $H\rightarrow$
diphotons and heavy-flavor dielectrons produced in $p+p$
collisions at 13 TeV. Different symbols represent different
particles in different $\eta$ ranges measured by the
ATLAS~\cite{30} and ALICE~\cite{31} Collaborations and scaled by
different factors marked in the panel. The dotted and dashed
curves are our fitted results by using Eqs. (8) and (10),
respectively. (b) The ratio of data to fit obtained from Eq. (8).
(c) The ratio of data to fit obtained from Eq. (10).}
\end{figure*}

{\section{Results and discussion}}

{\subsection{Comparison with data}}

Figure 1(a) shows the $p_{\rm T}$ spectra (the invariant
cross-sections), $Ed^3\sigma/dp^3$, of different hadrons with
given combinations and decay channels including $(\pi^++\pi^-)/2$
plus $\pi^0 \rightarrow \gamma\gamma$, $(K^++K^-)/2$ plus
$K_S^0\rightarrow\pi^0\pi^0$, $\eta \rightarrow \gamma\gamma$ plus
$\eta \rightarrow \pi^0\pi^+\pi^-$, $\omega\rightarrow e^+e^-$
plus $\omega\rightarrow \pi^0\pi^+\pi^-$ plus $\omega\rightarrow
\pi^0\gamma$, $(p+\bar p)/2$, $\eta'\rightarrow\eta\pi^+\pi^-$,
$\phi\rightarrow e^+e^-$ plus $\phi\rightarrow K^+K^-$,
$J/\psi\rightarrow e^+e^-$, and $\psi'\rightarrow e^+e^-$ produced
in $p+p$ collisions at 200 GeV. Different symbols represent
different particles and their different decay channels measured by
the PHENIX Collaboration~\cite{12} in the pseudorapidity range of
$|\eta|<0.35$. The results corresponding to $\pi$, $K$, $\eta$,
$\omega$, $p$, and $\eta'$ are re-scaled by multiplying by $10^6$,
$10^5$, $10^4$, $10^3$, $10^2$, and $10$ factors, respectively.
The results corresponding to $\phi$, $J/\psi$, and $\psi'$ are not
re-scaled.

In Fig. 1(a), the dotted and dashed curves are our fitted results
by using Eqs. (8) (for mesons and baryons) and (10) (for mesons)
or (12) (for baryons), respectively. The values of free parameters
($T$, $n$, and $a_0$), normalization constant ($\sigma_0$),
$\chi^2$, and number of degree of freedom (ndof) obtained from Eq.
(8) are listed in Table 1, while the values of parameters and
$\chi^2$/ndof obtained from Eqs. (10) or (12) are listed in Table
2. In Eq. (8), $m_0$ is taken to be the rest mass of $\pi$, $K$,
$\eta$, $\omega$, $p$, $\eta'$, $\phi$, $J/\psi$, and $\psi'$ for
the cases from $(\pi^++\pi^-)/2$ to $\psi'\longrightarrow e^+e^-$
sequenced according to the order shown in Fig. 1(a). In the fit
process at the quark level, the quark structure of $\pi^0$ results
in its $f(p_T)$ to be the half of the sum of $u\bar u$'s $f(p_T)$
and $d\bar d$'s $f(p_T)$. Because the constituent masses of $u$
and $d$ are the same~\cite{36}, $\pi^0$'s $f(p_T)$ is equal to
$u\bar u$'s $f(p_T)$ or $d\bar d$'s $f(p_T)$. The quark structure
of $\eta$ results in its $f(p_T)$ to be $\cos^2\phi \times u\bar
u$'s $f(p_T)$ + $\sin^2\phi \times s\bar s$'s $f(p_T)$ due to the
quark structures of $\eta_q$ and $\eta_s$, where
$\phi=39.3^{\circ}\pm1.0^{\circ}$ is the mixing angle~\cite{37}.
The quark structure of $\eta'$ results in its $f(p_T)$ to be
$\sin^2\phi \times u\bar u$'s $f(p_{\rm T})$ + $\cos^2\phi \times
s\bar s$'s $f(p_{\rm T})$.

To show departures of the fit from the data, following Fig. 1(a),
Figs. 1(b) and 1(c) show the ratios of data to fit obtained from
Eqs. (8) and (10) or (12), respectively. One can see that the fits
are around the data in the whole $p_T$ range, except for a few
sizeable departures. The experimental data for the mentioned
hadrons measured in $p+p$ collisions at 200 GeV by the PHENIX
Collaboration~\cite{12} can be fitted by Eqs. (8) (for mesons and
baryons) and (10) (for mesons) or (12) (for baryons). From the
values of $\chi^2$ and the data over fit ratio, one can see that
Eq. (10) or (12) can describe the data equally well as Eq. (8).

It seems that Eq. (10) or (12) is not necessary due to Eq. (8)
being good enough. In fact, the introduction of Eq. (10) or (12)
does not contain more parameters comparing with Eq. (8). Moreover,
Eq. (10) or (12) can tell more about the underlying physics than
Eq. (8). The effective temperature used in Eq. (10) or (12) is
related to the excitation degree of quark matter, while the
effective temperature in Eq. (8) is related to the excitation
degree of hadronic matter. In our opinion, Eqs. (10) and (12) are
necessary. We shall analyze sequentially the $p_{\rm T}$ spectra
of identified particles by using Eqs. (8) and (10) or (12) in the
following text.

\begin{table*}[!htb]
{\small Table 1. Values of $T$, $n$, $a_0$, $\sigma_0$, $\chi^2$,
and ndof corresponding to the dotted curves in Figs. 1(a), 3(a),
and 5(a) which are fitted by the TP-like function [Eq. (8)]. In
the case of ndof being less than 1, it appears as ``--" in the
table.} {\vspace{-0.4cm}} {\small\setlength{\tabcolsep}{1mm}
\begin{center}
\newcommand{\tabincell}[2]{\begin{tabular}{@{}#1@{}}#2\end{tabular}}
\begin{tabular} {ccccccccccccc}\\ \hline\hline
Figure & $y$ ($\eta$) & Particle & $T$ (GeV) & $n$ & $a_{0}$ & $\sigma_0$ (mb) & $\chi^2$/ndof \\
\hline
Fig. 1(a) & $|\eta|<0.35$ &\tabincell{c}{$(\pi^++\pi^-)/2$\\ $\pi^0$} & $0.129\pm0.001 $ & $9.449\pm0.020$ & $0.890\pm0.004$ & $37.044\pm0.348$ & 5/39\\
200 GeV   &               &\tabincell{c}{$(K^++K^-)/2$\\ $K^0_S$}     & $0.167\pm0.002 $ & $9.529\pm0.030$ & $1.027\pm0.004$ & $3.122\pm0.030$  & 7/27\\
          &               & $\eta$          & $0.195\pm0.002$ & $9.889\pm0.033$  & $1.000\pm0.003$ & $1.755\pm0.087$ & 6/32\\
          &               & $\omega$        & $0.193\pm0.001 $ & $9.460\pm0.100$  & $0.900\pm0.020$ & $3.073\pm0.030$ & 23/34\\
          &               & $(p+\bar{p})/2$ & $0.149\pm0.002 $ & $9.100\pm0.020$  & $1.040\pm0.003$ & $1.291\pm0.008$ & 11/13\\
          &               & $\eta'$         & $0.210\pm0.002 $ & $10.001\pm0.023$ & $0.980\pm0.003$ & $0.584\pm0.004$ & 4/8\\
          &               & $\phi$          & $0.245\pm0.002 $ & $10.559\pm0.023$ & $0.688\pm0.003$ & $0.334\pm0.003$ & 10/15\\
          &               & $J/\psi$        & $0.482\pm0.002 $ & $16.778\pm0.023$ & $0.901\pm0.004$ & $(5.320\pm0.132)\times10^{-4}$ & 4/22\\
          &               & $\psi'$         & $0.452\pm0.002 $ & $8.349\pm0.022$  & $0.959\pm0.003$ & $(9.234\pm0.008)\times10^{-5}$ & 1/--\\
\hline
Fig. 3(a) & $|\eta|<0.8$ & $\pi^++\pi^-$ & $0.130\pm0.001$ & $6.882\pm0.021$ & $ 0.937\pm0.002$ & $(3.961\pm0.032)\times10^{2}$ & 46/59\\
2.76 TeV  &              & $K^++K^-$     & $0.167\pm0.001 $ & $6.985\pm0.019$ & $1.209\pm0.003$ & $47.404\pm0.649$ & 32/54\\
          &              & $p+\bar{p}$   & $0.199\pm0.001 $ & $7.870\pm0.024$ & $1.064\pm0.003$ & $23.645\pm0.129$ & 56/45\\
          & $|y|<0.5$    & $\rho^0(770)$ & $0.205\pm0.001 $ & $6.987\pm0.021$ & $1.140\pm0.003$ & $14.167\pm0.069$ & 6/6\\
          &              & $\phi$        & $0.245\pm0.002 $ & $6.696\pm0.021$ & $1.010\pm0.004$ & $1.673\pm0.016$  & 7/17\\
          & $|y|<0.8$    & $\eta$        & $0.195\pm0.001 $ & $6.910\pm0.024$ & $1.023\pm0.003$ & $(3.340\pm0.064)\times10^{-10}$ & 7/7\\
          & $2.5<y<4$    & $J/\psi$      & $0.482\pm0.002 $ & $7.231\pm0.025$ & $1.225\pm0.003$ & $(2.181\pm0.029)\times10^{-2}$ & 4/3\\
\hline
Fig. 5(a) & $|y|<1$   &$(\pi^++\pi^-)/2$ & $0.129\pm0.001 $ & $4.862\pm0.021$ & $0.806\pm0.003$ & $(4.608\pm0.018)\times10^{2}$ & 57/18\\
13 TeV    &           &$(K^++K^-)/2$     & $0.167\pm0.001 $ & $6.179\pm0.018$ & $1.261\pm0.002$ & $(4.312\pm0.039)\times10^{1}$  & 3/13\\
          &           &$(p+\bar{p})/2$   & $0.199\pm0.002 $ & $4.768\pm0.023$ & $1.180\pm0.004$ & $(2.211\pm0.013)\times10^{1}$  & 8/22\\
          & $2<y<2.5$ &$J/\psi$ prompt   & $0.482\pm0.001 $ & $6.729\pm0.022$ & $1.581\pm0.003$ & $(1.097\pm0.001)\times10^{-2}$ & 115/10\\
          &           &$J/\psi$ from $b$     & $0.482\pm0.001 $ & $5.529\pm0.024$ & $1.877\pm0.002$ & $(1.850\pm0.020)\times10^{-3}$ & 35/10\\
          &           &$\psi(2s)$ prompt     & $0.578\pm0.001 $ & $7.603\pm0.021$ & $1.867\pm0.003$ & $(1.509\pm0.019)\times10^{-3}$ & 33/13\\
          &           &$\psi(2s)$ from $b$   & $0.578\pm0.001 $ & $5.989\pm0.023$ & $1.901\pm0.004$ & $(4.907\pm0.104)\times10^{-4}$ & 33/13\\
          &           &$D^0+\bar{D}^0$       & $0.497\pm0.001 $ & $6.624\pm0.022$ & $1.244\pm0.003$ & $(5.573\pm0.063)\times10^{-1}$ & 5/14\\
          &           &$D^{+}+D^-$           & $0.497\pm0.002 $ & $6.446\pm0.022$ & $1.231\pm0.005$ & $(2.805\pm0.059)\times10^{-1}$ & 11/13\\
          &           &$D^{*+}+D^{*-}$       & $0.497\pm0.001 $ & $6.566\pm0.023$ & $1.231\pm0.004$ & $(2.622\pm0.054)\times10^{-1}$ & 16/11\\
          &           &$D_{s}^{+}+D_{s}^{-}$ & $0.497\pm0.001 $ & $9.259\pm0.021$ & $2.217\pm0.004$ & $(7.876\pm0.149)\times10^{-2}$ & 5/12\\
\hline
\end{tabular}%
\end{center}}
\end{table*}

\begin{table*}[!htb]
{\small Table 2. Values of $T$, $n$, $a_0$, $\sigma_0$, $\chi^2$,
and ndof corresponding to the dashed curves in Figs. 1(a), 3(a),
and 5(a) which are fitted by the convolution [Eq. (10) or (12)] of
two or three TP-like functions. The quark structures are listed
together. In the case of ndof being less than 1, it appears as
``--" in the table.} {\vspace{-0.35cm}}
{\scriptsize\setlength{\tabcolsep}{1mm}
\begin{center}
\newcommand{\tabincell}[2]{\begin{tabular}{@{}#1@{}}#2\end{tabular}}
\begin{tabular} {ccccccccccccc}\\ \hline\hline
Figure & $y$ ($\eta$) & Particle & Quark structure & $T$ (GeV) & $n$ & $a_{0}$ & $\sigma_0$ (mb) & $\chi^2$/ndof \\
\hline
Fig. 1(a) & $|\eta|<0.35$ &\tabincell{c}{$(\pi^++\pi^-)/2$\\ $\pi^0$} & \tabincell{c}{$u\bar d$, $d\bar u$\\ $(u\bar u-d\bar d)/\sqrt{2}$} & $0.209\pm0.002$ & $7.774\pm0.025$ & $-0.540\pm0.020$ & $38.357\pm0.350$ & 6/39\\
200 GeV   &               &\tabincell{c}{$(K^++K^-)/2$\\ $K^0_S$} & \tabincell{c}{$u\bar s$, $s\bar u$\\ $d\bar s$}   & $0.196\pm0.001$ & $7.816\pm0.030$ & $-0.091\pm0.005$ & $2.913\pm0.026 $ & 4/27\\
          &               &$\eta$: $\eta_q$, $\eta_s$  & $(u\bar u+d\bar d)/\sqrt{2}$, $s\bar s$ & $0.212\pm0.001$ & $8.109\pm0.030$  & $0.000\pm0.004$  & $1.838\pm0.017$ & 4/32\\
          &               &$\omega$                    & $(u\bar u+d\bar d)/\sqrt{2}$            & $0.222\pm0.001$ & $8.394\pm0.013$  & $0.000\pm0.002$  & $2.854\pm0.026$ & 20/34\\
          &               &$(p+\bar{p})/2$             & $uud$, $\bar u\bar u\bar d$             & $0.162\pm0.002$ & $7.600\pm0.021$  & $-0.130\pm0.003$ & $1.282\pm0.011$ & 3/13\\
          &               &$\eta'$: $\eta_q$, $\eta_s$ & $(u\bar u+d\bar d)/\sqrt{2}$, $s\bar s$ & $0.233\pm0.002$ & $8.315\pm0.025$  & $0.000\pm0.003$  & $0.593\pm0.004$ & 3/8\\
          &               &$\phi$                      & $s\bar s$                               & $0.266\pm0.001$ & $9.022\pm0.022$  & $-0.107\pm0.003$ & $0.319\pm0.003$ & 10/15\\
          &               &$J/\psi$                    & $c\bar c$                               & $0.509\pm0.002$ & $14.545\pm0.025$ & $0.008\pm0.003$  & $(5.275\pm0.004)\times10^{-4}$ & 4/22 \\
          &               &$\psi'$                     & $c\bar c$                               & $0.503\pm0.002$ & $7.025\pm0.030$  & $0.055\pm0.004$  & $(9.232\pm0.008)\times10^{-5}$ & 1/-- \\
\hline
Fig. 3(a) & $|\eta|<0.8$ &$\pi^++\pi^-$  & $u\bar d$, $d\bar u$                & $0.209\pm0.001$ & $4.970\pm0.022$ & $-0.490\pm0.003$ & $(4.474\pm0.021)\times10^{2}$ & 67/59\\
2.76 TeV  &              &$K^++K^-$                  & $u\bar s$, $s\bar u$                    & $0.197\pm0.002$ & $5.223\pm0.017$ & $-0.058\pm0.004$ & $44.823\pm0.605 $ & 30/54\\
          &              &$p+\bar{p}$                & $uud$, $\bar u\bar u\bar d$             & $0.194\pm0.002$ & $5.867\pm0.021$ & $-0.120\pm0.003$ & $24.401\pm0.113$ & 31/45\\
          & $|y|<0.5$    &$\rho^0(770)$              & $(u\bar u+d\bar d)/\sqrt{2}$            & $0.252\pm0.001$ & $5.775\pm0.021$ & $-0.003\pm0.003$ & $14.514\pm0.075$ & 4/6\\
          &              &$\phi$                     & $s\bar s$                               & $0.266\pm0.001$ & $5.313\pm0.019$ & $0.018\pm0.001$  & $1.679\pm0.045$ & 6/17\\
          & $|y|<0.8$    &$\eta$: $\eta_q$, $\eta_s$ & $(u\bar u+d\bar d)/\sqrt{2}$, $s\bar s$ & $0.212\pm0.002$ & $5.341\pm0.022$ & $-0.021\pm0.003$ & $(3.441\pm0.021)\times10^{-10}$ & 10/7\\
          & $2.5<y<4$    &$J/\psi$                   & $c\bar c$                               & $0.509\pm0.002$ & $6.181\pm0.018$ & $0.155\pm0.004$  & $(2.179\pm0.007)\times10^{2}$ & 4/3 \\
\hline
Fig. 5(a) & $|y|<1$   &$(\pi^++\pi^-)/2$ & $u\bar d$, $d\bar u$ & $0.182\pm0.001$ & $4.401\pm0.024$ & $-0.390\pm0.002$ & $(4.579\pm0.016)\times10^{2}$  & 54/18\\
13 TeV    &           &$(K^++K^-)/2$         & $u\bar s$, $s\bar u$        & $0.196\pm0.001$ & $7.081\pm0.023$ & $-0.023\pm0.002$ & $(4.384\pm0.016)\times10^{1}$  & 1/13\\
          &           &$(p+\bar{p})/2$       & $uud$, $\bar u\bar u\bar d$ & $0.191\pm0.001$ & $3.832\pm0.021$ & $-0.090\pm0.003$ & $(2.207\pm0.011)\times10^{1}$  & 8/22\\
          & $2<y<2.5$ &$J/\psi$ prompt       & $c\bar c$                   & $0.509\pm0.001$ & $4.855\pm0.022$ & $0.291\pm0.003$  & $(1.056\pm0.002)\times10^{-2}$ & 97/10 \\
          &           &$J/\psi$ from $b$     & $c\bar c$                   & $0.509\pm0.001$ & $3.904\pm0.023$ & $0.447\pm0.002$  & $(1.812\pm0.016)\times10^{-3}$ & 28/10\\
          &           &$\psi(2s)$ prompt     & $c\bar c$                   & $0.675\pm0.002$ & $6.098\pm0.022$ & $0.435\pm0.003$  & $(1.490\pm0.017)\times10^{-3}$ & 21/13\\
          &           &$\psi(2s)$ from $b$   & $c\bar c$                   & $0.675\pm0.002$ & $4.531\pm0.019$ & $0.431\pm0.004$  & $(4.836\pm0.059)\times10^{-4}$ & 20/13\\
          &           &$D^0+\bar{D}^0$       & $c\bar u,\bar c u$          & $0.545\pm0.001$ & $5.125\pm0.017$ & $0.104\pm0.003$  & $(5.420\pm0.045)\times10^{-1}$ & 3/14\\
          &           &$D^{+}+D^-$           & $c\bar d,\bar c d$          & $0.545\pm0.002$ & $4.929\pm0.023$ & $0.098\pm0.004$  & $(2.718\pm0.043)\times10^{-1}$ & 8/13\\
          &           &$D^{*+}+D^{*-}$       & $c\bar d,\bar c d$          & $0.545\pm0.001$ & $5.025\pm0.019$ & $0.101\pm0.002$  & $(2.477\pm0.034)\times10^{-1}$ & 14/11\\
          &           &$D_{s}^{+}+D_{s}^{-}$ & $c\bar s,\bar c s$          & $0.545\pm0.001$ & $6.515\pm0.020$ & $0.549\pm0.003$  & $(7.406\pm0.099)\times10^{-2}$ & 5/12\\
\hline
\end{tabular}%
\end{center}}
\end{table*}

\begin{table*}[!htb]
{\small Table 3. Values of $T$, $n$, $a_0$, $\sigma_0$, $\chi^2$,
and ndof corresponding to the dotted curves in Figs. 2(a), 4(a),
and 6(a) which are fitted by the TP-like function [Eq. (8)].}
{\vspace{-0.35cm}} {\scriptsize\setlength{\tabcolsep}{1mm}
\begin{center}
\newcommand{\tabincell}[2]{\begin{tabular}{@{}#1@{}}#2\end{tabular}}
\begin{tabular} {ccccccccccccc}\\ \hline\hline
Figure & $y$ ($\eta$) & Particle & $T$ (GeV) & $n$ & $a_{0}$& $\sigma_0$ (mb) & $\chi^2$/ndof \\
\hline
Fig. 2(a) & $|\eta|<0.35$    &inclusive direct $\gamma$                    & $0.258\pm0.001$ & $9.413\pm0.020 $ & $1.750\pm0.004$ & $(4.836\pm0.044)\times10^{-3}$  & 2/14\\
200 GeV   &                  &$(e^++e^-)/2$                                & $0.155\pm0.002$ & $8.460\pm0.030 $ & $0.652\pm0.003$ & $(1.105\pm0.009)\times10^{-2}$  & 8/24\\
          & $1.4<|y|<2.0$    &\tabincell{c}{$(\mu^++\mu^-)/2$\\(open heavy decays)}& $0.125\pm0.001$ & $9.308\pm0.022 $ & $0.799\pm0.003$ & $(2.343\pm0.015)\times10^{-2}$ & 7/9\\
          & $1.2<|\eta|<2.2$ &Drell--Yan $\longrightarrow \mu^+\mu^-$      & $0.349\pm0.002$ & $8.849\pm0.023 $ & $2.200\pm0.004$ & $(1.559\pm0.001)\times10^{-7}$  & 8/8\\
          &                  &$c\bar c \longrightarrow \mu^+\mu^-$         & $0.385\pm0.002$ & $13.983\pm0.023$ & $1.509\pm0.003$ & $(4.227\pm0.004)\times10^{-9}$  & 10/11\\
          &                  &$b\bar b \longrightarrow \mu^{\pm}\mu^{\pm}$ & $0.445\pm0.002$ & $20.501\pm0.050$ & $2.260\pm0.030$ & $(6.917\pm0.006)\times10^{-12}$ & 8/6\\
\hline
Fig. 4(a) & $|y|<0.9$ &inclusive $\gamma$                        & $0.166\pm0.001$ & $6.791\pm0.020$ & $0.068\pm0.002$  & $(3.565\pm0.035)\times10^{2}$ & 32/14\\
2.76 TeV  & $2.5<y<4$ &\tabincell{c}{$\mu$\\(from heavy decays)} & $0.345\pm0.001$ & $7.528\pm0.021$ & $0.000\pm0.003$  & $1.480\pm0.002$ & 7/12\\
          & $|y|<0.8$ &\tabincell{c}{$e$\\(from beauty decays)}  & $0.315\pm0.001$ & $6.094\pm0.016$ & $1.000\pm0.004$  & $7.686\pm0.051$ & 3/16\\
          &           &\tabincell{c}{$e$\\(from heavy decays)}   & $0.165\pm0.001$ & $4.305\pm0.020$ & $-0.043\pm0.004$ & $(3.701\pm0.063)\times10^{2}$ & 7/21\\
          &           &\tabincell{c}{$(e^++e^-)/2$\\(inclusive)} & $0.155\pm0.002$ & $5.554\pm0.019$ & $-0.05\pm0.002$  & $2.726\pm0.057$ & 7/15\\
\hline
Fig. 6(a) & $|\eta|<1.37$ &$H\rightarrow$ diphoton & $0.150\pm0.001$ & $14.681\pm0.022 $ & $12.257\pm0.004$ & $(5.295\pm0.186)\times10^{-11}$ & 16/9\\
13 TeV    & $|\eta|<0.8$  &heavy dielectron        & $0.125\pm0.001$ & $8.811\pm0.019 $  & $2.281\pm0.003$  & $(7.581\pm0.034)\times10^{-1}$  & 6/13\\
\hline
\end{tabular}%
\end{center}}
\end{table*}

\begin{table*}[!htb]
{\small Table 4. Values of $T$, $n$, $a_0$, $\sigma_0$, $\chi^2$,
and ndof corresponding to the dashed curves in Figs. 2(a), 4(a),
and 6(a) which are fitted by the convolution [Eq. (10)] of two
TP-like functions. The participant quarks are listed together.}
{\vspace{-0.35cm}} {\scriptsize\setlength{\tabcolsep}{1mm}
\begin{center}
\newcommand{\tabincell}[2]{\begin{tabular}{@{}#1@{}}#2\end{tabular}}
\begin{tabular} {ccccccccccccc}\\ \hline\hline
Figure & $y$ ($\eta$) & Particle & Quark & $T$ (GeV) & $n$ & $a_{0}$& $\sigma_0$ (mb) & $\chi^2$/ndof \\
\hline
Fig. 2(a) & $|\eta|<0.35$    &inclusive direct $\gamma$                    & $u\bar u$ & $0.383\pm0.001$ & $6.793\pm0.024$ & $0.060\pm0.002$ & $(4.967\pm0.044)\times10^{-3}$  & 2/14\\
200 GeV   &                  &$(e^++e^-)/2$                                & $u\bar u$ & $0.236\pm0.002$ & $6.408\pm0.020$ & $-0.596\pm0.003$& $(1.192\pm0.008)\times10^{-2}$ & 5/24\\
          & $1.4<|y|<2.0$    &\tabincell{c}{$(\mu^++\mu^-)/2$\\(open heavy decays)}& $uc$ & $0.167\pm0.001$ & $6.035\pm0.025 $ & $-0.802\pm0.003$ & $(2.226\pm0.014)\times10^{-2}$ & 4/9\\
          & $1.2<|\eta|<2.2$ &Drell--Yan $\longrightarrow \mu^+\mu^-$      & $u\bar u$ & $0.418\pm0.002$ & $5.616\pm0.023$ & $0.398\pm0.004$ & $(1.571\pm0.001)\times10^{-7}$  & 8/8\\
          &                  &$c\bar c \longrightarrow \mu^+\mu^-$         & $c\bar c$ & $0.207\pm0.002$ & $4.072\pm0.025$ & $0.005\pm0.004$ & $( 4.206\pm0.004)\times10^{-9}$ & 8/11\\
          &                  &$b\bar b \longrightarrow \mu^{\pm}\mu^{\pm}$ & $b\bar b$ & $0.207\pm0.002$ & $5.653\pm0.024$ & $0.049\pm0.004$ & $(7.047\pm0.006)\times10^{-12}$ & 3/6\\
\hline
Fig. 4(a) & $|y|<0.9$ &inclusive $\gamma$                               & $u\bar u$ & $0.233\pm0.001$ & $5.383\pm0.019 $ & $-0.700\pm0.003$ & $(3.345\pm0.020)\times10^{2}$ & 27/14\\
2.76 TeV  & $2.5<y<4$ &\tabincell{c}{$\mu$\\(from heavy decays)}        & $c\bar c$ & $0.309\pm0.001$ & $4.554\pm0.022 $ & $-0.704\pm0.003$ & $1.364\pm0.003$ & 9/12\\
          & $|y|<0.8$ &\tabincell{c}{$e$\\(from beauty decays)}         & $b\bar b$ & $0.080\pm0.001$ & $1.993\pm0.018 $ & $-0.150\pm0.004$ & $7.323\pm0.042$ & 4/16\\
          &           &\tabincell{c}{$e$\\(from heavy decays)}          & $c\bar c$ & $0.206\pm0.002$ & $2.441\pm0.017 $ & $-0.894\pm0.002$ & $(3.720\pm0.043)\times10^{2}$ & 5/21\\
          &           &\tabincell{c}{$(e^++e^-)/2$\\(inclusive)}        & $u\bar u$ & $0.166\pm0.001$ & $3.724\pm0.018 $ & $-0.700\pm0.003$ & $2.770\pm0.021$ & 12/15\\
\hline
Fig. 6(a) & $|\eta|<1.37$ &$H\rightarrow$ diphoton & $c\bar c$ & $0.702\pm0.001$ & $4.573\pm0.019$  & $2.550\pm0.003$ & $(5.376\pm0.209)\times10^{-1}$ & 17/13\\
13 TeV    & $|\eta|<0.8$  & heavy dielectron       & $c\bar c$ & $0.166\pm0.001$ & $3.128\pm0.021$ & $-0.520\pm0.002$ & $(7.581\pm0.034)\times10^{-1}$ & 6/9\\
\hline
\end{tabular}%
\end{center}}
\end{table*}

\begin{figure*}[!htb]
\begin{center}
\includegraphics[width=15.0cm]{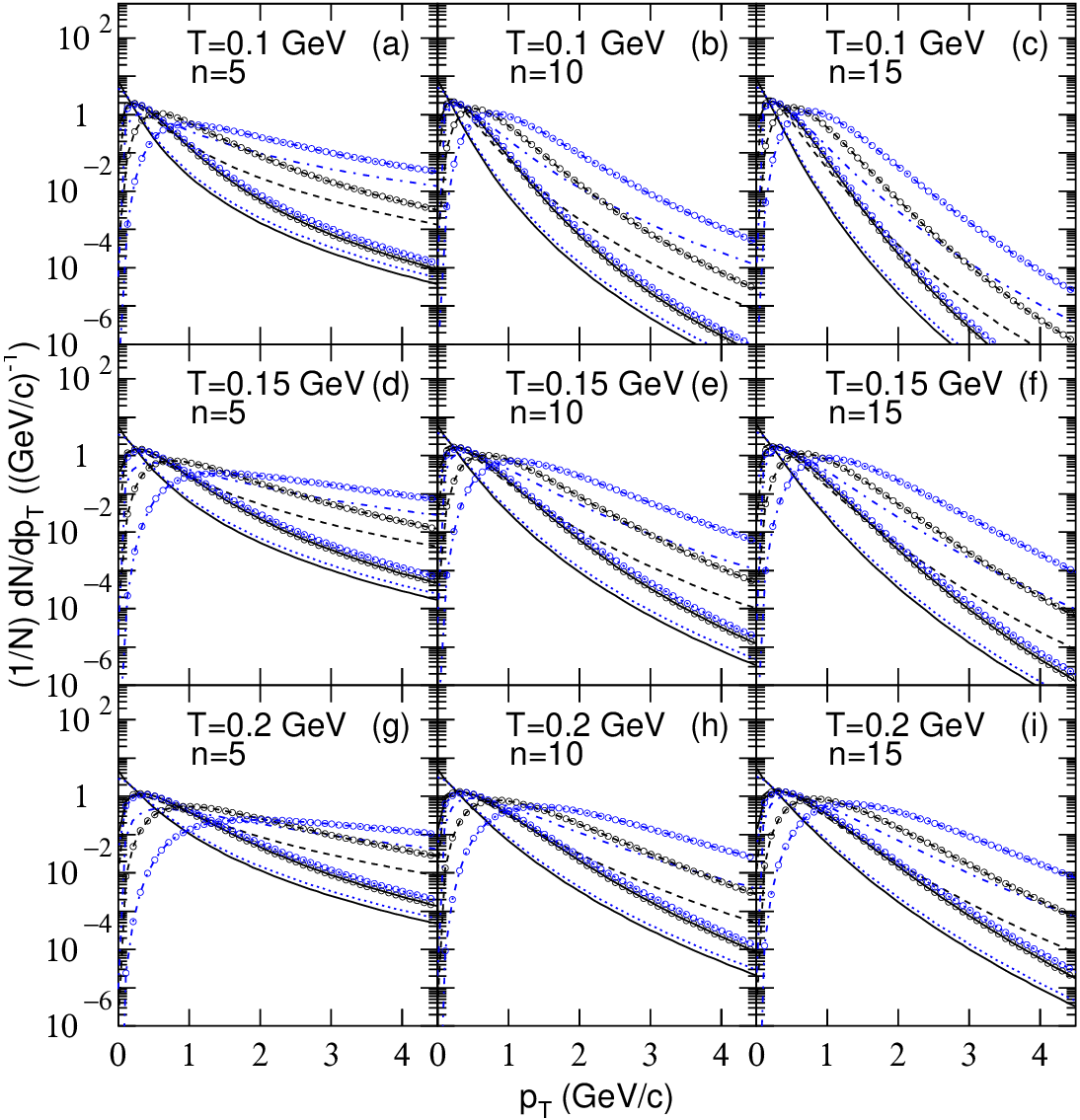}
\end{center}
{\small Fig. 7. Various pion spectra with different parameters in
Eqs. (8) and (10). From the upper panel [Figs. 7(a), 7(b), and
7(c)] to the middle panel [Figs. 7(d), 7(e), and 7(f)] and then to
the lower panel [Figs. 7(g), 7(h), and 7(i)], $T$ changes from 0.1
GeV to 0.15 GeV and then to 0.2 GeV. From the left panel to the
middle panel and then to the right panel, $n$ changes from 5 to 10
and then to 15. In each panel, the solid, dotted, dashed, and
dot-dashed curves without (with) open circles are obtained by
$a_0=-0.1$, 0, 1, and 2, respectively, from Eq. (8) [Eq. (10)].}
\end{figure*}

Figure 2(a) shows the invariant cross-sections of inclusive direct
photons and different leptons with given combinations and
production channels including $(e^++e^-)/2$, $(\mu^++\mu^-)/2$
(open heavy-flavor decays), Drell--Yan $\longrightarrow
\mu^+\mu^-$, $c\bar c \longrightarrow \mu^+\mu^-$, and $b\bar b
\longrightarrow \mu^{\pm}\mu^{\pm}$ produced in $p+p$ collisions
at 200 GeV. Different symbols represent different particles and
their production channels measured by the PHENIX
Collaboration~\cite{13,14,15,16} in different $\eta$ or $y$
ranges. The dotted and dashed curves are our fitted results by
using Eqs. (8) and (10), respectively, where two participant
quarks are considered in the formation of the mentioned particles.
The values of parameters and $\chi^2$/ndof obtained from Eqs. (8)
and (10) are listed in Tables 3 and 4, respectively. In Eq. (8),
$m_0$ is taken to be the rest mass of $\gamma$, $e$, $\mu$,
$2\mu$, $2\mu$, and $4\mu$ for the cases from inclusive direct
$\gamma$ to $b\bar b\longrightarrow \mu^{\pm}\mu^{\pm}$ sequenced
according to the order shown in Fig. 2(a), where $2\mu$ is two
times due to the continued two $2\mu$-related channels. In Eq.
(10), $m_{01}+m_{02}$ are taken to be the constituent masses of
$u+u$, $u+u$, $u+c$, $u+u$, $c+c$, and $b+b$ sequenced according
to the same order as particles.

Following Fig. 2(a), Figs. 2(b) and 2(c) show the ratios of data
to fit obtained from Eqs. (8) and (10), respectively. One can see
that the fits of the data are rather good in the whole $p_T$
range, except for a few sizeable departures. The experimental data
on the mentioned photons and leptons measured in $p+p$ collisions
at 200 GeV by the PHENIX Collaboration~\cite{13,14,15,16} can also
be fitted by Eqs. (8) and (10). From the values of $\chi^2$ and
the data over fit ratio, one can see that Eq. (10) can describe
the data equally well as Eq. (8).

Similarly to Fig. 1(a), Fig. 3(a) shows the invariant
cross-sections of various hadrons produced in $p+p$ collisions at
2.76 TeV. Different symbols represent different particles measured
by the ALICE Collaboration~\cite{17,18,19,20,21} in different
$\eta$ or $y$ ranges. The values of parameters and $\chi^2$/ndof
are listed in Table 1. The fit of $\rho$ at the quark level is the
same with $\pi^0$. Other particles and corresponding quarks are
discussed in Fig. 1(a). Similarly, Fig. 3(b) and 3(c) show the
ratios of data to fit obtained from Eqs. (8) and (10) or (12),
respectively. One can see that the fits of the data are rather
good in the whole $p_{\rm T}$ range, except for a few sizeable
departures. The experimental data on the mentioned hadrons
measured in $p+p$ collisions at 2.76 TeV by the ALICE
Collaboration~\cite{17,18,19,20,21} can be fitted by Eqs. (8) and
(10) or (12). From the values of $\chi^2$ and the data over fit
ratio, one can see that Eq. (10) or (12) can describe the data
equally well as Eq. (8).

Similarly to Fig. 2(a), Fig. 4(a) shows the invariant
cross-sections of photons and different leptons with given
combinations and production channels including inclusive $\gamma$,
$\mu$ from heavy-flavor hadron decays, $e$ from beauty hadron
decays, $e$ from heavy-flavor hadron decays, and inclusive
$(e^++e^-)/2$ produced in $p+p$ collisions at 2.76 TeV. Different
symbols represent different particles measured by the ALICE
Collaboration~\cite{22,23,24,25} in different $y$ ranges. The
values of parameters and $\chi^2$/ndof are listed in Table 2. In
Eq. (8), $m_0$ is taken to be the rest mass of $\gamma$, $\mu$,
$e$, $e$, and $e$ for the cases from inclusive $\gamma$ to
inclusive $(e^++e^-)/2$ sequenced according to the order shown in
Fig. 4(a), where $e$ is three times due to the continued three
$e$-related channels. In Eq. (10), $m_{01}+m_{02}$ are taken to be
the constituent masses of $u+u$, $c+c$, $b+b$, $c+c$, and $u+u$
sequenced according to the same order as particles. Following Fig.
4(a), Figs. 4(b) and 4(c) show the ratios of data to fit obtained
from Eqs. (8) and (10), respectively. One can see that the fits
agree with the data in the whole $p_{\rm T}$ range, except for a
few departures. The experimental data on the mentioned photons and
leptons measured in $p+p$ collisions at 2.76 TeV by the ALICE
Collaboration~\cite{22,23,24,25} can be fitted by Eqs. (8) and
(10). From the values of $\chi^2$ and the data over fit ratio, one
can see that Eq. (10) can describe the data equally well as Eq.
(8).

Similarly to Figs. 1(a) and 3(a), Fig. 5(a) shows the invariant
cross-sections of different hadrons produced in $p+p$ collisions
at 13 TeV. Different symbols represent different particles
measured by the CMS~\cite{26} or LHCb~\cite{27,28,29}
Collaborations in different $y$ ranges. The values of parameters
and $\chi^2$/ndof are listed in Table 1. Except for the first five
groups of particles and corresponding quarks which are discussed
in Fig. 1(a), $m_0$ in Eq. (8) for other particles is taken to be
the rest mass of $\psi(2s)$, $\psi(2s)$, $D^0$, $D^+$, $D^{*+}$,
and $D^+_s$ for the cases from $\psi(2s)$ prompt to $D^+_s+D^-_s$
sequenced according to the order shown in Fig. 5(a) from the left
to right panels. Meanwhile, $m_{01}+m_{02}$ in Eq. (10) for other
cases are taken to be the constituent masses of $c+c$, $c+c$,
$c+u$, $c+d$, $c+d$, and $c+s$ sequenced according to the same
order as particles. Following Fig. 5(a), Figs. 5(b) and 5(c) show
the ratio of data to fit obtained from Eqs. (8) and (10) or (12),
respectively. One can see that the fits close to the data in the
whole $p_{\rm T}$ range, except for a few departures. The
experimental data on the mentioned hadrons measured in $p+p$
collisions at 13 TeV by the CMS~\cite{26} and LHCb~\cite{27,28,29}
Collaborations can be fitted by Eqs. (8) and (10) or (12). From
the values of $\chi^2$ and the data over fit ratio, one can see
that Eq. (10) or (12) can describe the data equally well as Eq.
(8).

Similarly to Figs. 2(a) and 4(a), Fig. 6(a) shows the invariant
cross-sections of $H\rightarrow$ diphotons and heavy flavor
dielectrons produced in $p+p$ collisions at 13 TeV. Different
symbols represent different particles measured by the
ATLAS~\cite{30} or ALICE~\cite{31} Collaborations in different
$\eta$ ranges. The values of parameters and $\chi^2$/ndof are
listed in Table 2. In Eq. (8), $m_0$ is taken to be the rest
masses of $2\gamma$ ($=0$) and $2e$ sequenced according to the
order shown in Fig. 6(a). In Eq. (10), both types of particles
correspond to the same $m_{01}+ m_{02}$, i.e. the constituent
masses of $c+c$. Following Fig. 6(a), Figs. 6(b) and 6(c) show the
ratios of data to fit obtained from Eqs. (8) and (10),
respectively. One can see that the fits agree with the data in the
whole $p_{\rm T}$ range, except for a few departures. The
experimental data of diphotons and dielectrons measured in $p+p$
collisions at 13 TeV by the ATLAS~\cite{30} and ALICE~\cite{31}
Collaborations can be fitted by Eqs. (8) and (10). From the values
of $\chi^2$ and the data over fit ratio, one can see that Eq. (10)
can describe the data equally well as Eq. (8).
\\

{\subsection{Discussion on parameters}}

We now analyze the tendencies of the free parameters. The values
of effective temperature $T$ for the emissions of different
hadrons do not depend on collision energy. This situation is
different for the emissions of photons and leptons, in which there
is a clear dependence on energy. This reflects that the emission
processes of photons and leptons are more complex than those of
hadrons. In central (pseudo)rapidity region, $T$ shows an
incremental tendency with the increase of particle or quark mass.
This is understandable that more collision energies are deposited
to produce massive hadrons or to drive massive quarks to take part
in the process of photon and lepton production. In the
forward/backward (pseudo)rapidity region, $T$ is expected to be
less than that in central (pseudo)rapidity region due to less
energy deposited.

The values of power index $n$ are very large with small
fluctuations in this study. In the Tsallis
statistics~\cite{6,7,7a,7b,8,9,10,11}, $n=1/(q-1)$, where $q$ is
an entropy index that characterizes the degree of equilibrium or
non-equilibrium. Generally, $q=1$ corresponds to an equilibrium
state. A larger $q$ than 1 corresponds to a non-equilibrium state.
This study renders that the values of $q$ are very close to 1,
which means that the collision system considered by us is
approximately in an equilibrium state. The functions based on
statistical methods are applicable in this study. In particular,
with the increasing collision energy, $n$ decreases and then $q$
increases slightly. This means that the collision system gets
further away from the equilibrium state at higher energy.

The values of revised index $a_0$, for the fits in Figs. 1(a) and
3(a), listed in Table 1 show that maybe Eq. (8) is not useful
because $a_0 \approx 1$. However, the values of $a_0$ listed in
Table 3 show that Eq. (8) is indeed necessary because $a_0 \neq
1$. The values of $a_0$ for the fits in Fig. 5(a) and listed in
Table 1 are larger than 1 for nearly all heavy-flavor particles,
while the values of $a_0$ for others are around 1. The values of
$a_0$, for the fits in Figs. 2(a), 4(a), and 6(a), listed in
Tables 3 and 4 are not equal to 1 in most cases. In general, Eq.
(8) is necessary in the data-driven analysis because $a_0 \neq 1$
in most cases. In fact, Tables 1--4 show specific $a_0$ and
corresponding collision energy, (pseudo)rapidity range, and
particle type. Strictly, there are only two cases with $a_0=1$,
that is the meson $\eta$ production in $pp$ collisions with
$|\eta|<0.35$ at 200 GeV (Table 1) and electron $e$ from beauty
decays in $pp$ collisions with $|y|<0.8$ at 2.76 TeV (Table 3).

To see the dependences of the spectra on free parameters, Figure 7
presents various pion spectra with different parameters in Eqs.
(8) and (10). From the upper panel [Figs. 7(a), 7(b), and 7(c)] to
the middle panel [Figs. 7(d), 7(e), and 7(f)] and then to the
lower panel [Figs. 7(g), 7(h), and 7(i)], $T$ changes from 0.1 GeV
to 0.15 GeV and then to 0.2 GeV. From the left panel to the middle
panel and then to the right panel, $n$ changes from 5 to 10 and
then to 15. In each panel, the solid, dotted, dashed, and
dot-dashed curves without (with) open circles correspond to the
spectra with $a_0=-0.1$, 0, 1, and 2, respectively, from Eq. (8)
[Eq. (10)]. One can see that the probability in high $p_T$ region
increases with increasing $T$, decreases with increasing $n$, and
increases with increasing $a_0$. From negative to positive, $a_0$
determines the shape in the low-$p_{\rm T}$ region.

From the shapes of curves in Fig. 7, one can see that the
parameter $a_0$ introduced in the TP-like function [Eq. (8)] by us
determines mainly the trend of curve in low-$p_{\rm T}$ region. If
the production of light particles via resonance decay affect
obviously the shape of spectrum, one may use a more negative $a_0$
in the fit. If the decay or absorbtion effect of heavy particles
in hot and dense medium in participant region affect obviously the
shape of spectrum, one may use a more positive $a_0$ in the fit.
Due to the introduction of $a_0$, the TP-like function is more
flexible than the Tsallis--Pareto-type function. In fact, $a_0$ is
a sensitive quantity to describe the influence of the production
of light particles via resonance decay and the decay or absorbtion
effect of heavy particles in hot and dense medium. Indeed, the
introduction of $a_0$ is significant.

Before summary and conclusions, we would like to point out that
ref.~\cite{7b} proposes an alternative form of parametrization for
the Tsallis-like function which also well describes the spectra in
the low-$p_{\rm T}$ region, which we give as a major improvement
of our fit. Indeed, although many theoretical or modelling works
are proposed in high energy collisions, more works with different
ideas are needed as the ways to systemize the experimental data in
the field with fast progress.
\\

{\section{Summary and Conclusions}}

We summarize here our main observations and conclusions.

\begin{enumerate}

\item{The transverse momentum spectra in terms of the (invariant)
cross-section of various particles (different hadrons with given
combinations and decay channels, photons, and different leptons
with given combinations and production channels) produced in high
energy proton-proton collisions have been studied by a TP-like
function (a revised Tsallis--Pareto-type function). Meanwhile, the
transverse momentum spectra have also been studied by a new
description in the framework of participant quark model or the
multisource model at the quark level. In the model, the source
itself is exactly the participant quark. Each participant quark
contributes to the transverse momentum spectrum to be the TP-like
function.}

\item{For a hadron, the participant quarks are in fact constituent
quarks. The transverse momentum spectrum of the hadron is the
convolution of two or more TP-like functions. For a photon or
lepton, the transverse momentum spectrum is the convolution of two
TP-like functions due to two participant quarks, e.g. projectile
and target quarks, taking part in the collisions. The TP-like
function and the convolution of a few TP-like functions can fit
the experimental data of various particles produced in
proton-proton collisions at 200 GeV, 2.76 TeV, and 13 TeV measured
by the PHENIX, ALICE, CMS, LHCb, and ATLAS Collaborations.}

\item{The values of effective temperature for the emissions of
different hadrons do not depend on collision energy, while for the
emissions of photons and leptons there is an obvious dependence on
collision energy. This reflects the fact that the emission
processes of photons and leptons are more complex than those of
hadrons. In central (pseudo)rapidity region, the effective
temperature shows an increasing tendency with the increase of
particle or quark mass. This reflects the fact that more collision
energy is deposited to produce massive hadrons or to drive massive
quarks to take part in the process of photon and lepton
production.}

\item{The values of power index are very large, which means that
the values of entropy index are very close to 1. The collision
system considered in this study is approximately in an equilibrium
state. The functions based on statistical methods are applicable
in this study. In particular, with the increase of collision
energy, the power index decreases and then the entropy index
increases slightly. This means that the collision system gets
further away from the equilibrium state at higher energy, though
the entropy index is still close to 1 at the LHC.}

\item{The values of revised index show that the TP-like function
is indeed necessary due to the fact that this index is not equal
to 1. In the TP-like function and its convolution, the effective
temperature, power index, and revised index are sensitive to the
spectra. In various pion spectra from the TP-like function and its
convolution of two, the probability in high transverse momentum
region increases with the increase of effective temperature,
decreases with the increase of power index, and increases with the
increase of revised index. From negative to positive, the revised
index determines the shape in low transverse momentum region,
which is sensitive to the contribution of resonance decays.}
\\

\end{enumerate}

\vskip0.5cm

\noindent {\bf Data Availability}

The data used to support the findings of this study are included
within the article and are cited at relevant places within the
text as references.
\\
\\
\\
{\bf Ethical Approval}

The authors declare that they are in compliance with ethical
standards regarding the content of this paper.
\\
\\
\\
{\bf Disclosure}

The funding agencies have no role in the design of the study; in
the collection, analysis, or interpretation of the data; in the
writing of the manuscript; or in the decision to publish the
results.
\\
\\
\\
{\bf Conflict of Interest}

The authors declare that there are no conflicts of interest
regarding the publication of this paper.
\\
\\
\\

\noindent {\bf Acknowledgements}

The first author (P.P.Y.) thanks Prof. Dr. David Blaschke and his
colleagues of Bogoliubov Laboratory for Theoretical Physics of
Joint Institute for Nuclear Research (Russia) for their
hospitality, in where this work was partly performed. Her work was
supported by the China Scholarship Council (Chinese Government
Scholarship) under Grant No. 202008140170 and the Shanxi
Provincial Innovative Foundation for Graduate Education under
Grant No. 2019SY053. The work of the second author (F.H.L.) was
supported by the National Natural Science Foundation of China
under Grant Nos. 11575103 and 11947418, the Scientific and
Technological Innovation Programs of Higher Education Institutions
in Shanxi (STIP) under Grant No. 201802017, the Shanxi Provincial
Natural Science Foundation under Grant No. 201901D111043, and the
Fund for Shanxi ``1331 Project" Key Subjects Construction. The
work of the third author (R.S.) was supported by the financial
supports from ALICE Project No. SR/MF/PS-01/2014-IITI(G) of
Department of Science \& Technology, Government of India.
\\

\end{multicols}
\end{document}